# THE PROJECT-X INJECTOR EXPERIMENT: A NOVEL HIGH PERFORMANCE FRONT-END FOR A FUTURE HIGH POWER PROTON FACILITY AT FERMILAB*

S. Nagaitsev#, S. Holmes, D. Johnson, M. Kaducak, R. Kephart, V. Lebedev, S. Mishra, A. Shemyakin, N. Solyak, R. Stanek, V. Yakovlev, Fermilab, Batavia, IL 60510, USA
P. Ostroumov, ANL, Argonne, IL 60439, USA
D. Li, LBNL, Berkeley, CA 94720, USA
P. Singh, M. Pande, S. Malhotra, Bhabha Atomic Research Centre, Mumbai-400085, India

*Abstract*

A multi-MW proton facility, Project X, has been proposed and is currently under development at Fermilab. We are carrying out a program of research and development aimed at integrated systems testing of critical components comprising the front end of Project X. This program, known as the Project X Injector Experiment (PXIE), is being undertaken as a key component of the larger Project X R&D program. The successful completion of this program will validate the concept for the Project X front end, thereby minimizing a primary technical risk element within Project X. PXIE is currently under construction at Fermilab and will be completed over the period FY12-17. PXIE will include an H- ion source, a CW 2.1-MeV RFQ and two superconductive RF (SRF) cryomodules providing up to 25 MeV energy gain at an average beam current of 1 mA (upgradable to 2 mA). Successful systems testing will also demonstrate the viability of novel front end technologies that are expected find applications beyond Project X.

## PXIE GOALS AND PLAN

A unique feature of Project X [1] is the capability of delivering a MW-range proton beam to several experiments quasi-simultaneously with a beam structure that can be adjusted to each experiment's needs. In the suggested scheme, this is achieved as follows:

(a) prepare a "nearly CW" beam, in which many RF buckets may be empty but the beam current, averaged over several μs, is constant. The undesired bunches are removed in the Medium Energy Beam Transport (MEBT) by a chopping system. This quasi-CW beam has a repetitive time structure (RS).

(b) accelerate the prepared beam. The optimum solution requires acceleration by SRF cavities after MEB at 2.1 MeV;

(c) distribute the beam to experimental stations by an RF separator at 1 GeV and 3 GeV.

The RF separation of bunches successfully supports multiple experiments operation at CEBAF [2]. The acceleration with SRF at low energies has been used for a long time at ANL [3] (at much lower average currents) and recently is being commissioned at SARAF [4], but has not been demonstrated at parameters required for the Project X. The wideband chopper, needed to prepare bunch patterns, is a unique device currently beyond state of the art.

The PXIE program [5] will address both (a) and (b) with the specific goals of demonstrating effective bunch-by-bunch chopping of the CW beam delivered from the RFQ and efficient acceleration of the 1 mA RS beam with minimal emittance dilution through at least 15 MeV.

PXIE consists of an H- ion source (IS); a Low Energy Beam Transport (LEBT); a CW 2.1-MeV RFQ; a MEBT; two SRF cryomodules (CM) operating at 2K - a Half Wave Resonator (HWR) and a Single Spoke Resonator (SSR1); a High Energy Beam Transport (HEBT); and a beam dump (Fig.1).

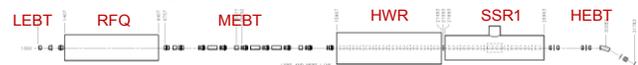

Figure 1: PXIE layout.

The concept of PXIE subsystems was described in [5]. The end-to-end simulations can be found in [6]. Main beam parameters are shown in Table 1.

Table 1: Nominal beam parameters at the exit of each section.

|  | IS/LEBT | RFQ | MEBT | HWR | SSR1 |
|---|---|---|---|---|---|
| Energy, MeV | 0.03 | 2.1 | 2.1 | 11 | 25 |
| Average current, mA | 5 | 5 | 1 | 1 | 1 |
| Time structure | DC | CW | RS | RS | RS |
| $\varepsilon_\perp$, μm |  |  |  |  | < 0.25 |
| $\varepsilon_\parallel$, keV-ns |  |  |  |  | ≤ 1 |

\* $\varepsilon_\perp$ is the rms normalized transverse emittance, and $\varepsilon_\parallel$ is the rms longitudinal emittance.

The PXIE design and construction is being carried out by a collaboration between Fermilab, ANL, LBNL, SLAC, SNS, and Indian laboratories in a staged approach.

___________________________________________
*Work supported by the U.S. Department of Energy under contract No. DE-AC02-07CH11359
#nsergei@fnal.gov

The beam is planned to pass through the LEBT in FY14, through the RFQ in FY15, and full PXIE parameters are expected to be demonstrated by the end of 2018.

## STATUS OF SUBSYSTEMS

The ion source has been procured from D-Pace Inc. [7], was characterized and used in the beam dynamics studies at LBNL, and presently is being assembled at Fermilab.

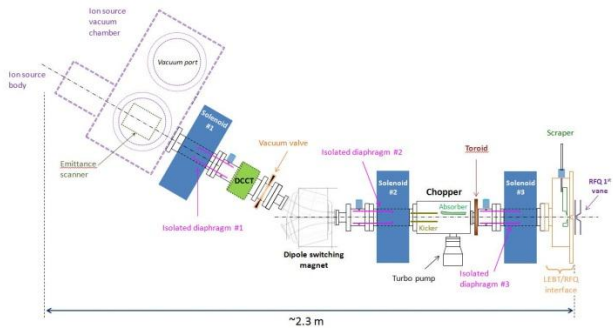

Figure 2: Conceptual design of the LEBT.

The LEBT has been conceptually designed (Fig.2). The 2.2 m distance between the ion source and the RFQ entrance ensures a low hydrogen flux from the ion source into the RFQ. Because the design of the Project X front end includes two ion sources for redundancy, the final PXIE assembly will include a combining dipole magnet though only one ion source will be used. Focusing is provided by three solenoids. A chopping system will be used for several purposes: to generate a pulse mode during commissioning, in a normal operation, to shut down injection into the RFQ for machine protection, possibly, to interrupt the beam for the residual ion removal from the MEBT, and, if found necessary, to assist the MEBT chopper.

One of the concerns is a growth of effective emittance due to neutralization following chopper pulses. To address this, the LEBT will be assembled first in a simple, straight configuration, in which the beam can be neutralized either along the entire line or only upstream of the chopper. In the latter case, the large distance to the IS allows effective clearing of the residual ions by combination of a positive potential on a stopping electrode and a constant transverse electric field. The ion removal solves the issue with temporal changes of the beam envelope but increases the emittance growth due to space charge [6]. Experiments will show which mode is preferable. Presently, the LEBT elements are at various stages of designing, procurement, and assembly.

The 4-vane, 4.4 m 162.5 MHz RFQ (Fig. 3) has been designed and the fabrication has started in August, 2013 at LBNL. The detailed description of the RFQ design and status can be found in Ref. [8], and the design of the coupler is described in Ref. [9]. The complete assembly of the RFQ is planned to be delivered to Fermilab at the end of 2014.

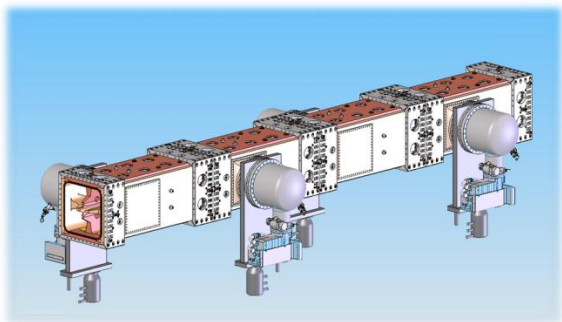

Figure 3: The RFQ conceptual design [9].

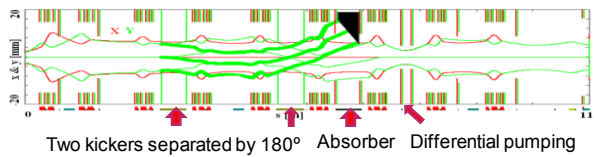

Figure 4: Scheme of MEBT optics and the beam envelope. The thin lines are the central trajectory and 3σ envelope ($\varepsilon_{rms\_n}$=0.25 mm mrad) of the passing beam, and the thick lines are the Y envelope of the chopped-out beam. Red- quadrupoles, blue- bunching cavities.

The design of the ~10 m MEBT was presented in [10], and the beam trajectory is shown in Fig. 4. Its most challenging part is the chopping system that forms the required bunch pattern by deflecting the undesired bunches of the initially CW beam into an absorber. This deflection is made by two broadband, travelling-wave kickers separated by 180º of the betatron phase advance and operating synchronously. The electric length of each kicker is 0.5 m, the plate separation is 16 mm, and the plate voltage with respect to ground is ±250V. The plates have opposite polarity voltages for bunch passing and removal. Presently two versions of the kicker, distinguished by the characteristic impedance of the kicker structure (50 and 200 Ohm), are being developed.

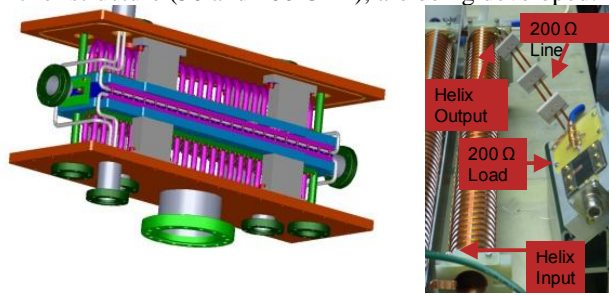

Figure 5: The MEBT kicker. A 3D model of the 50-Ohm kicker (left). Part of the vacuum box is removed for the presentation purpose. A model of the 200 Ohm kicker (right).

In the 50 Ohm version, the beam is deflected by the voltage applied to planar electrodes connected in vacuum by coaxial cables with the length providing the necessary delays (Fig. 5). The kicker prototype is being assembled.

RF and thermal properties of an 8-plate model with electrodes and cable in the final version were successfully tested. The kicker will be powered by a commercially available RF-amplifier with a specially shaped driving signal [11].

The 200 Ohm kicker [12] consists of two helixes wound around grounded cylinders (Fig. 5). Presently it is at the stage of the pre-prototype fabrication.

The 200 Ohm kicker will be driven by broadband, DC coupled GaN switches in push-pull configuration being developed at Fermilab [12]. Operation of a single switch up to 500 V was demonstrated, and recently a fully functional 100V driver was successfully tested.

The kickers displace the undesired bunches by ~6$\sigma_y$ to the absorber surface. The absorber is specified to withstand 21 kW, twice the full nominal beam power, focused into a spot with $\sigma_{x,y}$ ~2mm. A preliminary conceptual design is complete, and ¼-size prototype has been successfully tested with an electron beam (Fig. 6). Preliminary results can be found in Ref. [13].

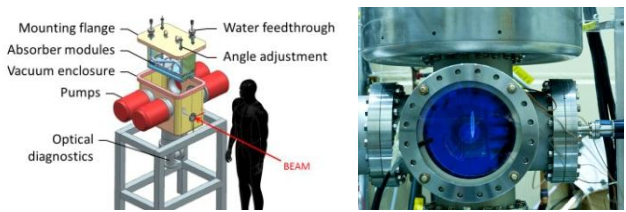

Figure 6: Absorber development: a conceptual design (left). Right – the OTR image of a 5.5-kW, 9-mm electron beam (bright ellipse) on the surface of the absorber prototype.

Transverse and longitudinal focusing in MEBT will be provided by 25 quadrupoles combined into doublets and triplets, and by three 162.5-MHz bunching cavities, correspondingly. The cavities are at the final stage of mechanical design at Fermilab. The quadrupoles and dipole correctors have been designed at BARC in India, and the prototypes are being tested.

The PXIE SRF cryomodules are being fabricated at ANL (HWR) and at Fermilab (SSR1). The HWR cryomodule [14] contains 8 162.5-MHz, nominally 1.7 MV, $\beta = 0.11$ cavities and 8 SC solenoids, with built-in dipole correctors and a BPM attached to each solenoid. Design of cavities and the cryomodule is complete (Fig. 7). Prototypes of the RF coupler and solenoid with dipole correctors have been build and successfully tested. Nb parts for all cavities have been fabricated, and two cavities will be tested in FY14.

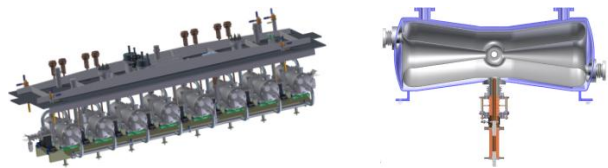

Figure 7: HWR CM design: a cavity-solenoid string attached to the top lid (left), a cavity with a coupler (right).

The SSR1 (Fig. 8) cryomodule consists of eight 325-MHz single-spoke resonators ($\beta = 0.22$) and 4 solenoids. Similar to the HWR CM, all solenoids will have corrector coils and BPMs. The conceptual design of the cryomodule is complete, and most of elements are at the stage of technical design.

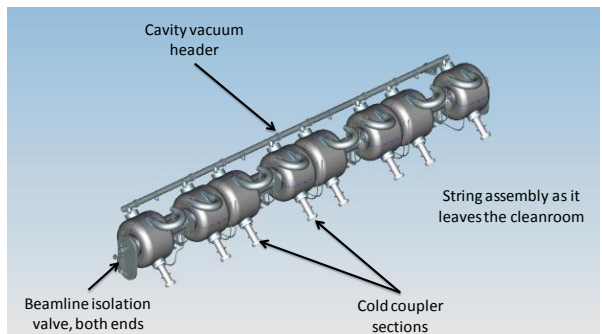

Figure 8: The SSR1 cavity and solenoid string assembly.

Details of the SSR1 cavity development is reported in [15]. The bare cavity and the He vessel are shown in Figure 9. The first SSR1 cavity prototype was manufactured by ZANON, the cavity was tested at Fermilab's Vertical Test Stand (VTS) and outfitted with a helium vessel. The second prototype was manufactured by ROARK, tested and showed acceptable performance. Subsequently, ten other SSR1 cavities were manufactured and delivered to Fermilab.

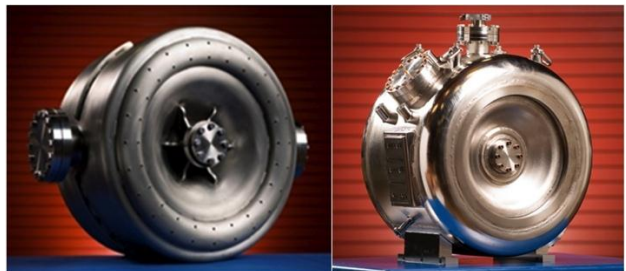

Figure 9: SSR1 cavity, bare (left) and dressed (right).

The SSR1 is followed by the HEBT that accommodates the beam diagnostics to measure the beam properties and the beam extinction for RF buckets emptied by the MEBT chopper. Its design is at the conceptual stage (Fig. 10) as well as is the PXIE beam dump at the end of the beam line.

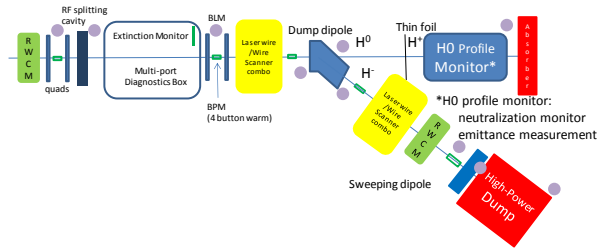

Figure 10: The HEBT scheme.

The PXIE will be located in the existing Cryomodule Test Facility (CMTF) building which will also include an enclosure for ILC and Project X cryomodules testing. A cryo-plant supporting operation of both installations, which will have 500 W cooling power at 2 K, is under construction. The upstream portion of the PXIE cave, long enough to accommodate the warm portion of PXIE, has been assembled with concrete shielding blocks. The infrastructure, controls, and diagnostics are at various stages of design and implementation.